\documentclass[12pt,onecolumn]{IEEEtran}
\usepackage{graphicx}
\usepackage{epsfig,amsfonts,amsmath,amssymb,amstext,amsthm}
\usepackage{threeparttable}
\usepackage{epstopdf}
\usepackage{footnote}
\usepackage{setspace}

\newcommand{\bm}[1]{\mbox{\pmb{$#1$}}}

\newcommand{\dref}[1]{(\ref{#1})}
\newcommand{\Tr}{\mbox{Tr}}

\newtheorem {Lemma}{Lemma}

\begin{document}

\title{\Huge{Performance Analysis of Dual-User Macrodiversity MIMO Systems with Linear Receivers
in Flat Rayleigh Fading}}

\author{Dushyantha A. Basnayaka,~\IEEEmembership{Student Member,~IEEE,}
        Peter J. Smith,~\IEEEmembership{Senior Member,~IEEE}
        and~Philippa A. Martin,~\IEEEmembership{Senior Member,~IEEE}
\thanks{D. A. Basnayaka, P. J. Smith and P. A. Martin are with the Department of Electrical and Computer Engineering, University of Canterbury, Christchurch, New
Zealand. E-mail:\{dush, p.smith, p.martin\}@elec.canterbury.ac.nz.}
\thanks{D. A. Basnayaka is supported by a University of Canterbury International Doctoral Scholarship.}}
%
%

%
\maketitle

\begin{abstract}
The performance of linear receivers in the presence of co-channel
interference in Rayleigh channels is a fundamental problem in
wireless communications. Performance evaluation for these systems is
well-known for receive arrays where the antennas are close enough to
experience equal average SNRs from a source. In contrast, almost no
analytical results are available for macrodiversity systems where
both the sources and receive antennas are widely separated. Here,
receive antennas experience unequal average SNRs from a source and a
single receive antenna receives a different average SNR from each
source. Although this is an extremely difficult problem, progress is
possible for the two-user scenario. In this paper, we derive closed
form results for the probability density function (pdf) and
cumulative distribution function (cdf) of the output signal to
interference plus noise ratio (SINR) and signal to noise ratio (SNR)
of minimum mean squared error (MMSE) and zero forcing (ZF) receivers
in independent Rayleigh channels with arbitrary numbers of receive
antennas. The results are verified by Monte Carlo simulations and
high SNR approximations are also derived. The results enable further
system analysis such as the evaluation of outage probability, bit
error rate (BER) and capacity.
\end{abstract}


\section{Introduction}\label{sec:introduction}
A macrodiversity multiple input multiple output (MIMO) system is
considered in this paper to denote a system where both the transmit
antennas and receive antennas are widely separated. As a result, the
slow fading experienced on all links is different and each link has
a different average signal to noise ratio (SNR). There is
considerable interest in such systems from a variety of
perspectives. They arise naturally in network MIMO
\cite{Wang10,Yu09,Siva07} and in other types of base station (BS)
collaboration \cite{Sander09}. A simplified model, namely the
classical circular Wyner model \cite{Wy94}, is widely used in
network MIMO systems, but is too restrictive to be useful in modern
macrodiversity MIMO systems such as those proposed in 3GPP
LTE-Advanced standards \cite{SaKiMo10}. In the circular Wyner model,
it is assumed that a given cell only experiences interference from
two adjacent cells and this interference has a fixed level given by
a particular fraction of the desired signal power. In contrast, the
general model discussed in this paper does not make any such
assumptions and assumes as many interferers as the system permits
with arbitrary powers \cite{Sander09}. Macro-diversity can also
occur as a result of collaborative MIMO ideas \cite[p.~69]{Big07}.
In addition, the fundamental channel model, where each link has a
different SNR, is closely related to the MIMO channel models in
\cite{Werner06}. Note that the multiple distributed transmit
antennas could correspond to many single-antenna users, one MIMO
transmitter with distributed
antennas or variations of the two.\\
The performance of linear receivers in such a macrodiversity system
has been investigated via simulation \cite{Mari09,Lin09}, but very
few analytical results appear to be available. Hence, in this paper
we consider an analytical treatment of signal to interference plus
noise (SINR) performance for two types of linear receivers: minimum
mean squared error (MMSE) and zero forcing (ZF) receivers. We focus
on the baseline case of a flat fading Rayleigh channel, where the
links are all independent but have different SNRs due to the
geographical separation. This subject has been well-studied in the
micro-diversity case \cite{Gao98,Peter07,Winters94, Gore02} where
there may be distributed sources, but the receive antennas are
closely spaced. The performance metric of interest is the SINR/SNR
distribution since this also leads to results for bit-error-rate
(BER), symbol-error-rate (SER), outage probability and capacity,
etc.\\
The difficulty in analyzing macrodiversity systems is that there is
no coherent methodology currently available to handle the type of
channel matrices that occur. In independently distributed Rayleigh
channels, basic results in statistics have been used to great effect
\cite{Gao98, Peter07}. In the presence of correlation, if a
Kronecker correlation structure is assumed, there are also many
results available \cite{Salz94, Shiu00, Shin06, Mckay07}. These
results tend to have their origins in multivariate statistics
\cite{Muir82} and make heavy use of hypergeometric function theory
\cite{Gross89}. Unfortunately, no such theory seems to be available
for complex Gaussian matrices where every element has a different
variance, i.e., the macrodiversity case. As a result, we focus on a
case where progress is possible; the two-user scenario. For this
scenario, the sources are two single antenna users or a single user
with two widely separated transmit antennas. The sources communicate
with an arbitrary number of base stations each with a single receive
antenna or a single base station with an arbitrary number of widely
distributed receive antennas. In this scenario, user 1 is detected
with user 2 as the interferer and then vice-versa. A particular
example of this scenario is also considered, where a three sector
cluster in a network MIMO system communicates with two single
antenna users. For the general two-user scenario, we are able to
derive the exact closed form SINR/SNR distribution for both MMSE and
ZF receivers. In addition to the exact SINR/SNR analysis, we also
derive high SNR results for the SER of MMSE and ZF receivers. These
results lead to a simple metric which relates system performance to
the average link SNRs and therefore provides insight into the effect
of these SNRs. In particular, we establish the following key
observations and results. An exact analysis for the dual-user case
is obtained. The methodology is not extendable to the multi-user
case which suggests that approximations and/or bounds may be needed
to handle more than two users. Simple, high SNR approximations are
provided for the SER. These novel expressions provide a functional
link between performance and the channel powers which is more
accurate than previous measures such as orthogonality. The analysis
shows that the performance of MMSE and ZF receivers becomes more
different when the users have "parallel" channel powers. At low
SINR, performance is enhanced by diversity (when the desired user
has roughly equal channel powers at the receive antennas), whereas
at high SINR performance is enhanced by a subset of receive antennas
having a high desired power and a low interference power.\\
The rest of the paper is laid out as follows. Section II describes
the system model and receiver types. The main analysis is given in
Sec. III. Sections IV and V give numerical results and conclusions,
respectively.

\section{System Model and Receiver Types}\label{sec:system_model}
\subsection{System Model}\label{subsec:model}
\begin{figure}[!t]
\centering
\includegraphics[scale=0.7]{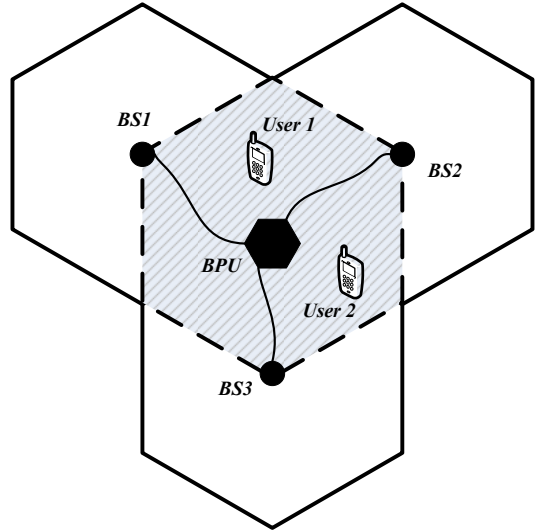}
\noindent \caption{A network MIMO system with a 3 sector cluster.}
\label{sec1:fig:macro}
\end{figure}

Consider two single-antenna users communicating with $n_R$
distributed receive antennas in an independent flat Rayleigh fading
environment. The $\mathcal{C}^{n_R \times 1}$ received vector is
given by
\vspace{0mm}
\begin{eqnarray}
\bm{r} &=& \bm{Hs} + \bm{n},
\end{eqnarray}
\noindent where $\bm{n}$ is the $\mathcal{C}^{n_R \times 1}$
additive-white-Gaussian-noise (AWGN) vector, $\bm{s}=\left ( s_1 ,
s_2 \right )^T$ contains the two transmitted symbols from user 1 and
user 2 and $\bm{H}$ is the $\mathcal{C}^{n_R \times 2}$ channel
matrix. The complex transmit vector, \bm{s}, is normalized so that
$E \left \{ |s_1|^2 \right \}= E \left \{ |s_2|^2 \right \}=1$. The
Gaussian noise vector, $\bm{n} \sim \mathcal{CN}\left ( \bm{0},
\sigma^2 \bm{I} \right )$, has independent entries with $E \left \{
|n_i|^2 \right \}= \sigma^2$, for $i=1,2, \dots, n_R$. The channel
matrix contains independent elements, $\bm{H}_{ik} \sim
\mathcal{CN}\left ( 0 , P_{ik} \right )$, where $E \left \{
|\bm{H}_{ik}|^2 \right \}=
P_{ik}$. \\
A particular example of this scenario is shown in
Fig.\ref{sec1:fig:macro}, where three BSs collaborate via a central
backhaul processing unit (BPU) in the shaded three sector cluster to
serve two single antenna users. In Fig. \ref{sec1:fig:macro}, it is
clear that the geographical spread of users and receivers gives rise
to a $3 \times 2$ channel matrix, $\bm{H}$, where all the $P_{ik}$
values are different.
\subsection{Receiver Types}\label{subsec:types}
At the receiver, the $n_R$ distributed antennas perform linear
combining. Hence, the output of the combiner is $\tilde{\bm{r}} =
\bm{W}^H \bm{r}$, where $\bm{W}$ is an $n_R \times 2$ weight matrix.
The form of $\bm{W}$ and the resulting output SINR/SNR are well
known for MMSE and ZF receivers. These results are summarized below.
Without loss of generality, we assume that the index of the desired
user is $i=1$ and we denote the first column of $\bm{W}$ by
$\bm{w}_1$. In practice, user 1 is the desired user with user 2 the
interferer, followed by user 2 being detected in the presence of
interference from user 1. From \cite{Gao98, Winters94, Jiang11} the
combining vector and output SINR of the MMSE receiver are given by
\vspace*{0mm}
\begin{equation}
\bm{w}_1 = \bm{R}^{-1} \bm{h}_1,
\end{equation}
\vspace{2mm}
\begin{equation}
\mbox{SINR} = \bm{h}_1^H \bm{R}^{-1} \bm{h}_1,
\label{eq:mmse:sinr_corrected}
\end{equation}

\noindent where $\bm{R} = \bm{h}_2 \bm{h}_2^H + \sigma^2 \bm{I}$ and
$\bm{h}_1, \bm{h}_2$ denote columns 1 and 2 of $\bm{H}$. From
\cite{Winters94, Gore02} the combining matrix $\bm{W} = \bm{H} \left
( \bm{H}^H \bm{H} \right )^{-1}$ and output SNR of the ZF receiver
for $n_R \geq 2$ are given by
\begin{equation}
\mbox{SNR} = \frac{1}{\sigma^2 \left[ \left ( \bm{H}^H \bm{H} \right
)^{-1} \right]_{11}}, \label{eq:zf:snr}
\end{equation}
\noindent where $\left[\bm{B}\right]_{11}$ indicates the $\left(1,1
\right)^{th}$ element of matrix $\bm{B}$.

\section{System Analysis}\label{sec:system_analysis}
In this section, we derive the CDFs of the output SINR/SNR of MMSE
and ZF receivers. First, we present some useful results as follows.

\subsection{ZF Analysis} \label{subsec:ZF_analysis}
Let $\tilde{Z}$ be the output SNR of a ZF receiver as given in
\dref{eq:zf:snr}. $\tilde{Z}$ can be written as \cite{Jiang11}
\begin{align}
\tilde{Z} &= \frac{1}{\sigma^2} \bm{h}_1^H \bm{M} \bm{h}_1,
\label{eq:zf:snr1}
\end{align}
where $\bm{M}=\bm{I} - \bm{h}_2 \left( \bm{h}_2^H \bm{h}_2
\right)^{-1} \bm{h}^H_2$. The characteristic function (cf) of
$\tilde{Z}$ is \cite{Muir82,Proakis01}\vspace{0mm}
\begin{eqnarray}
\phi_{\tilde{Z}}(t) &=& E \left \{ e^{jt\tilde{Z}} \right \} = E
\left \{ e^{\frac{jt}{\sigma^2} \bm{h}_1^H \bm{M} \bm{h}_1 }
\right\}. \label{eq:zf:cf1}
\end{eqnarray}
Note that $\bm{M}$ and $\bm{h}_1$ are independent and the pdfs of
$\bm{h}_1$ and $\bm{h}_2$ are given by
\vspace{0mm}
\begin{equation}
f(\bm{h}_k) = \frac{1}{\pi^{n_R} \left|\bm{P}_k \right|}
e^{-\bm{h}_k^H \bm{P}_k^{-1} \bm{h}_k }, \hspace{3mm} \mbox{for}
\hspace{3mm} k=1,2, \label{eq:pdfs}
\end{equation}
\noindent where the $n_R \times n_R$ matrix $\bm{P}_k =
\mbox{diag}(P_{1k}, P_{2,k}, \dots , P_{n_R,k})$. Next, by using
Lemma \ref{lemma:complex_gaussian}, the cf conditioned on $\bm{h}_2$
becomes \vspace{0mm}
\begin{equation}
\phi_{\tilde{Z}}(t | \bm{h}_2) = \frac{1}{\left| \bm{I} - jt
\frac{1}{\sigma^2}\bm{M} \bm{P}_1 \right|}. \label{eq:zf:cf2}
\end{equation}
Substituting $\bm{M}$ in \dref{eq:zf:cf2} and rearranging gives
\begin{align}
\phi_{\tilde{Z}}(t | \bm{h}_2) = \frac{\bm{h}^H_2 \bm{h}_2}{\left|
\bm{D} \right| \left ( \bm{h}^H_2 \bm{D}^{-1} \bm{h}_2 \right ) },
\label{eq:zf:cf3}
\end{align}
\noindent where $\bm{D}= \bm{I} - \frac{1}{\sigma^2}jt\bm{P}_1$. In
Appendix \ref{app:A0}, the full cf is obtained by averaging the
conditional cf in \dref{eq:zf:cf3} over $\bm{h}_2$. Then, the
resulting cf is inverted to give the final result, the cdf in
\dref{eq:zf:cdf5}. In \dref{eq:zf:cdf5}, the coefficients,
$\tilde{\varphi}_{ik}$, $\tilde{\psi}_{ik}$ and
$\tilde{\omega}_{ik}$ are defined in \dref{eq:zf:three_constants}
and the arguments, $\tilde{n}_{ik}$, $\tilde{\alpha}_{ik}$,
$\tilde{m}_{ik}$ and $\tilde{\beta}_{ik}$ are defined in
\dref{eq:zf:basic_constants0}. The integrals,
$\tilde{I}_1\left(.\right)$, $\tilde{I}_2\left(.\right)$ and
$\tilde{I}_3\left(.\right)$ are given in
\dref{eq:zf:double_integral1}-\dref{eq:zf:double_integral3}.
\begin{table*}\small
\begin{align}
F_{\tilde{Z}}(z) = \sigma^{2} \sum_{i=1}^{n_R} \sum_{k\neq i}^{n_R}
\tilde{\varphi}_{ik}\tilde{I}_1\left(\tilde{n}_{ik},\tilde{\alpha}_i,
\tilde{m}_{ik},\tilde{\beta}_i,z \right) +
\tilde{\psi}_{ik}\tilde{I}_2\left(\tilde{n}_{ik},\tilde{\alpha}_i,
\tilde{m}_{ik},\tilde{\beta}_i,z \right) +
\tilde{\omega}_{ik}\tilde{I}_3\left(\tilde{n}_{ik},\tilde{\alpha}_i,
\tilde{m}_{ik},\tilde{\beta}_i,z \right). \label{eq:zf:cdf5}
\end{align}
\end{table*}
\begin{table*}\small
\begin{align}
\tilde{I}_1\left(a,b,c,d,x \right) = \frac{1}{bc -ad} \left[ \ln
\left( \frac{bc}{ad} \right) -  e^{-bx} e^{\frac{adx}{c}} E_1 \left
( \frac{adx}{c} \right ) +  E_1 \left ( bx \right ) \right],
\label{eq:zf:I1:solution}
\end{align}
\vspace{4pt}
\begin{align}
\tilde{I}_2\left(a,b,c,d,x \right) = \left[ \frac{dx}{c \left( bc
-ad \right) } +  \frac{d}{\left( bc -ad \right)^2 } \right]e^{-bx}
e^{\frac{adx}{c}} E_1 \left ( \frac{adx}{c} \right ) -
\frac{d}{\left( bc -ad \right)^2 } E_1 \left ( bx \right ) - \frac{d
+ b\left( bc -ad \right)}{\left( bc
-ad \right)^2 } \ln \left( \frac{bc}{ad} \right) \nonumber \\
+ \frac{1 - e^{-bx}}{a \left( bc -ad \right) },
\label{eq:zf:I2:solution}
\end{align}
\vspace{4pt}
\begin{align}
\tilde{I}_3\left(a,b,c,d,x \right) = \left[ \frac{ax}{c \left( bc
-ad \right) } +  \frac{a}{\left( bc -ad \right)^2 } \right]e^{-bx}
e^{\frac{adx}{c}} E_1 \left ( \frac{adx}{c} \right ) -
\frac{a}{\left( bc -ad \right)^2 } E_1 \left ( bx \right ) -
\frac{a}{\left( bc -ad \right)^2 } \ln \left(
\frac{bc}{ad} \right)  \nonumber \\
+ \frac{a\left( 1 - e^{-bx} \right)}{bc\left( bc -ad \right) } +
\frac{1 - e^{-bx}}{bcd}. \label{eq:zf:I3:solution}
\end{align}
\hrulefill \vspace{4pt}\\
\end{table*}
\subsection{MMSE Analysis} \label{subsec:MMSE_analysis}
The complete MMSE analysis can be found in \cite{Dush11}. In this
paper, we repeat a few of the initial steps which are also needed
for the later high SNR results. Let Z be the output SINR of an MMSE
receiver given by \dref{eq:mmse:sinr_corrected}. The cf of Z is
given by \cite{Gao98,Proakis01}\vspace{0mm}
\begin{eqnarray}
\phi_{Z}(t) &=& E \left \{ e^{jtZ} \right \} = E \left \{ e^{jt
\bm{h}_1^H \bm{R}^{-1} \bm{h}_1 } \right\}. \label{eq:mmse:cf1}
\end{eqnarray}
As in the ZF analysis, we first obtain the cf conditioned on
$\bm{h}_2$. The conditional cf is given by
\begin{equation}
\phi_{Z}(t | \bm{h}_2) = \frac{1}{\left| \bm{I} - jt \bm{R}^{-1}
\bm{P}_1 \right|}. \label{eq:mmse:cf2}
\end{equation}
Substituting $\bm{R}$ in \dref{eq:mmse:cf2} and rearranging gives
\begin{equation}
\phi_{Z}(t | \bm{h}_2) = \frac{\sigma^2 + \bm{h}_2^H \bm{h}_2
}{\left|\bm{D} \right| \left ( \sigma^2 + \bm{h}_2^H \bm{D}^{-1}
\bm{h}_2 \right )}. \label{eq:mmse:cf3}
\end{equation}
Comparing $\dref{eq:zf:cf3}$ with $\dref{eq:mmse:cf3}$ we observe
that, as expected, the MMSE results converge to the ZF results as
$\sigma^2\rightarrow 0$. Again, Lemma
\ref{lemma:expected_value_of_ratios} can be used to average the
conditional cf in \dref{eq:mmse:cf3} to obtain the unconditional cf
and in turn the cdf of the output SINR of an MMSE receiver as in
\cite{Dush11}. For the sake of completeness, we give the final
results in \dref{eq:mmse:cdf5}-\dref{eq:mmse:I3:solution}.
\begin{table*}
\small
\begin{align}
F_Z(z) = \sigma^{2} \sum_{i=1}^{n_R} \sum_{k\neq i}^{n_R}
\varphi_{ik}I_1\left( \sigma^2
\tilde{n}_{ik},\tilde{\alpha}_i,\tilde{m}_{ik},\frac{\tilde{\beta}_i}{\sigma^2},z
\right) + \psi_{ik}I_2\left(\sigma^2
\tilde{n}_{ik},\tilde{\alpha}_i,\tilde{m}_{ik},\frac{\tilde{\beta}_i}{\sigma^2},z
\right) + \omega_{ik}I_3\left(\sigma^2
\tilde{n}_{ik},\tilde{\alpha}_i,\tilde{m}_{ik},\frac{\tilde{\beta}_i}{\sigma^2},z
\right). \label{eq:mmse:cdf5}
\end{align}
\vspace{4pt}
\begin{align}
I_1\left(a,b,c,d,x \right) &= \frac{1}{bc - ad} \left [
e^{\frac{a}{c}} E_1 \left ( \frac{a}{c} \right) - e^{\frac{b}{d}}
E_1 \left ( \frac{b}{d} \right) +  e^{\frac{b}{d}} E_1 \left (
\frac{b}{d} + bx \right) - e^{-bx}e^{\frac{\left( 1 + dx
\right)a}{c}}  E_1 \left (\frac{\left( 1 + dx \right)a}{c} \right)
\right]. \label{eq:mmse:I1:solution}
\end{align}
\vspace{4pt}
\begin{align}
I_2\left(a,b,c,d,x \right) = \left[ \frac{d}{\left( bc-ad\right)^2}
+ \frac{1 + dx }{c\left( bc-ad\right)} \right]e^{-bx}e^{\frac{\left(
1 + dx \right)a}{c}} E_1 \left (\frac{\left( 1 + dx \right)a}{c}
\right) - \left. \frac{cd + bc - ad }{c\left( bc-ad\right)^2}\right.
e^{\frac{a}{c}}E_1 \left ( \frac{a}{c} \right) \nonumber \\
+ \frac{d}{\left( bc-ad\right)^2} e^{\frac{b}{d}} \left [ E_1 \left
( \frac{b}{d} \right) - E_1 \left ( \frac{b}{d} + bx \right) \right
] + \frac{1 - e^{-bx}}{a\left( bc-ad\right)}.
\label{eq:mmse:I2:solution}
\end{align}
\vspace{4pt}
\begin{align}
I_3\left(a,b,c,d,x \right) = \left[ \frac{a}{\left( bc-ad\right)^2}
+ \frac{ax }{c\left( bc-ad\right)} \right] e^{-bx}e^{\frac{\left( 1
+ dx \right)a}{c}} E_1 \left (\frac{\left( 1 + dx \right)a}{c}
\right) + \left. \frac{ \left. ad^2+abd - b^2c \right.}{d^2\left(
bc-ad\right)^2} \right. e^{\frac{b}{d}} \left [ E_1 \left (
\frac{b}{d} \right)
\right. \nonumber \\
\left. - E_1 \left ( \frac{b}{d} + bx \right) \right ] -
\frac{a}{\left( bc-ad\right)^2} e^{\frac{a}{c}} E_1 \left (
\frac{a}{c} \right) + \frac{1 - e^{-bx}}{d\left( bc-ad\right)}.
 \label{eq:mmse:I3:solution}
\end{align}
\hrulefill \vspace{4pt}\\
\footnotesize{$^*$Results pertaining to MMSE receiver}.
\end{table*}
The cdf is defined in \dref{eq:mmse:cdf5} in terms of
$I_1\left(.\right)$, $I_2\left(.\right)$ and $I_3\left(.\right)$
which are given in
\dref{eq:mmse:I1:solution}-\dref{eq:mmse:I3:solution}. Finally, the
necessary constants $\varphi_{ik}$, $\psi_{ik}$ and $\omega_{ik}$
are given by
\begin{subequations}
\begin{align}
\!\!\!\!\varphi_{ik} \!&=\! P_{i1}^{n_R-2}\!\left( \sigma^2 \tilde{\eta}_{ik} \!+\! \tilde{\Delta}_i \tilde{\eta}_{ik} \!+\! \tilde{\zeta}_{ik}-(n_R-2)P_{i2}\tilde{\eta}_{ik} \right),\\
%
%
%
\psi_{ik} &= \sigma^2 P_{i1}^{n_R-2} \tilde{\eta}_{ik} \tilde{\xi}_{ik},\\
\omega_{ik} &= -P_{i2}^2 P_{i1}^{n_R-3} \tilde{\eta}_{ik}.
\end{align}
\end{subequations}
%
%
%
%
\section{SER Approximations}\label{sec:high_snr_analysis}
In this section, we derive high SNR approximations for the SER of
MMSE and ZF receivers. This is motivated by the complexity of the
exact analysis and the importance of finding a simple, functional
link between performance and the average link SNRs.
\subsection{ZF Analysis} \label{subsec:high_snr_ZF}
The conditional cf in \dref{eq:zf:cf3} is a ratio of quadratic forms
in $\bm{h}_2$. Hence, $\phi_{\tilde{Z}}(t) = E \{ \phi_{\tilde{Z}}(t
| \bm{h}_2) \}$ is the mean of a ratio of quadratic forms which can
be approximated by the Laplace approximation \cite{Lib94} as
\begin{align}
\phi_{\tilde{Z}}(t) \approx \frac{E \{ \bm{h}^H_2 \bm{h}_2 \}
}{\left| \bm{D} \right| E \left \{ \bm{h}^H_2 \bm{D}^{-1} \bm{h}_2
\right \}} = \frac{\Tr \left(
 \bm{P}_2 \right)}{|\bm{D}| \Tr \left( \bm{D}^{-1} \bm{P}_2 \right)}. \label{eq:high_zf:cf0}
\end{align}
Note that the second equality in \dref{eq:high_zf:cf0} follows from
the result, $E \{\bm{u}^H \bm{Q} \bm{u} \} = \Tr \left( \bm{Q}
\right) $, where $\bm{u}\sim \mathcal{CN} \left ( 0, \bm{I} \right
)$ and $\bm{Q}$ is a Hermitian matrix. Expanding the denominator of
\dref{eq:high_zf:cf0} gives
\begin{align}
\phi_{\tilde{Z}}(t) \approx \frac{\Tr \left( \bm{P}_2 \right)
}{\sum_{i=1}^{n_R} P_{i2} \prod_{k\neq i}^{n_R} \left( 1 -
\frac{jt}{\sigma^2}P_{k1}\right)}, \label{eq:high_zf:cf1}
\end{align}
\noindent which follows since $\bm{D}$ and $\bm{P}_2$ are diagonal
and $\left| \bm{D} \right|$ is the product of the diagonal entries
of $\bm{D}$. As the SNR grows, $\sigma^2\rightarrow 0$ and keeping
only the dominant power of $\sigma^2$ in \dref{eq:high_zf:cf1} gives
\begin{align}
\phi_{\tilde{Z}}(t) \approx \left.\frac{ \Tr \left( \bm{P}_2
\right)}{\left|\bm{P}_1 \right| \Tr \left(\bm{P}_1^{-1} \bm{P}_2
\right)} \right. \frac{\sigma^{2\left( n_R - 1 \right)}}{\left(-jt
\right)^{n_R -1}}. \label{eq:high_zf:cf2}
\end{align}
Defining $\vartheta \left (\bm{P}_1, \bm{P}_2 \right)=\frac{\Tr
\left. \bm{P}_2 \right.}{\left |\bm{P}_1 \right| \Tr
\left(\bm{P}_1^{-1} \bm{P}_2 \right)}$ gives a metric which
encapsulates the effects of the power matrices $\bm{P}_1$ and
$\bm{P}_2$. For many modulations, the SER can be evaluated as a
single integral of the moment generating function of the SNR
\cite{SiAl00}. The mgf of the SNR is $\mathcal{M}_{\tilde{Z}}
\left(s\right) = \phi_{\tilde{Z}} \left( -js \right)$. As an
example, for MPSK the SER is \cite{SiAl00, Gold00}
\begin{align}
\tilde{P}_s = \frac{1}{\pi} \int_0^{T} \mathcal{M}_{\tilde{Z}}
\left( -\frac{g}{\sin^2 \theta} \right ) d\theta,
\label{eq:high_zf:ser0}
\end{align}
\noindent where $g=\sin^{2}\left(\pi / M \right)$ and
$T=\frac{\left(M-1\right)\pi}{M}$. Substituting
\dref{eq:high_zf:cf2} in \dref{eq:high_zf:ser0} gives the
approximation
\begin{align}
\tilde{P}_s \approx \left(\tilde{G}_a
\bar{\gamma}\right)^{-\tilde{G}_d}. \label{eq:high_zf:ser1}
\end{align}
\noindent In \dref{eq:high_zf:ser1}, the average SNR is
$\bar{\gamma}=\frac{1}{\sigma^{2}}$, and the diversity gain and
array gain are given by
%
$$
\tilde{G}_d =n_R - 1, \hspace{10mm} \tilde{G}_a = \left( \vartheta
\left (\bm{P}_1, \bm{P}_2 \right) \tilde{\mathcal{I}}
\right)^{-1/(n_R-1)}. \nonumber
$$
%
The constant integral, $\tilde{\mathcal{I}}$, is given by
\begin{align}
\tilde{\mathcal{I}} = \frac{1}{\pi} \int_0^{T} \left( \frac{
\sin^2\theta}{g} \right)^{\left. n_R-1\right.} d\theta.
\label{eq:high_zf:integral}
\end{align}
Note that the simple representation in \dref{eq:high_zf:ser1} shows
the diversity order of $n_R-1$ and the effect of the link powers on
array gain controlled by the metric $\vartheta \left (\bm{P}_1,
\bm{P}_2 \right)$. The integral $\tilde{\mathcal{I}}$ can be solved
in closed form and the final result is given in
\dref{eq:high_zf:integral_solution}.
\begin{table*}[!t]
\normalsize
\begin{align}
\tilde{\mathcal{I}} = \frac{1}{\pi g^{\left. n_R-1 \right.}} \left
\{ \frac{T}{2^{2\left( n_R-1 \right)}} \left(\frac{2n_R - 2}{n_R -
1} \right)  + \frac{(-1)^{n_R - 1 }}{2^{2n_R-3} } \sum_{k=0}^{n_R-2}
(-1)^{k} \left(\frac{2n_R - 2}{k} \right) \frac{\sin \left(
2\left(n_R - k - 1 \right)T\right)}{2\left(n_R -k -1 \right )}
\right \} , \label{eq:high_zf:integral_solution}
\end{align}
\hrulefill \vspace{0pt}\\
\end{table*}
\subsection{MMSE Analysis} \label{subsec:high_snr_MMSE}
By using the Laplace approximation for the expectation of the
conditional cf in \dref{eq:mmse:cf3}, we obtain
\begin{align}
\phi_{Z}(t) \approx \frac{\sigma^2 + \Tr \left(
 \bm{P}_2 \right)  }{|\bm{D}| \left( \sigma^2 + \Tr \left( \bm{D}^{-1} \bm{P}_2 \right) \right)}. \label{eq:high_mmse:cf0}
\end{align}
As the SNR grows, $\sigma^2 \rightarrow 0$ and keeping only the
dominant power of $\sigma^2$ in \dref{eq:high_mmse:cf0} gives
\begin{align}
\phi_{Z}(t) \approx \frac{\Tr \left( \bm{P}_2 \right)\left(
\frac{1}{-jt\bar{\gamma}}\right)^{n_R -1 }}{\left| \bm{P}_1 \right|
\left( \Tr \left(\bm{P}_1^{-1} \bm{P}_2 \right) -jt \right)}.
\label{eq:high_mmse:cf1}
\end{align}
As in the ZF analysis, the SER for MPSK can be approximated by
\begin{align}
P_s \approx \left(G_a \bar{\gamma}\right)^{-G_d},
\label{eq:high_mmse:ser1}
\end{align}
where the diversity gain and array gain are given by
\begin{align}
G_d &= n_R - 1, \hspace{15mm} G_a = \left(\vartheta \left (\bm{P}_1,
\bm{P}_2 \right) \mathcal{I}\left( \bm{P}_1, \bm{P}_2 \right)
\right)^{-1/(n_R-1)}. \label{eq:high_mmse:array_gain}
\end{align}
\noindent In \dref{eq:high_mmse:array_gain}, $\mathcal{I}\left(
\bm{P}_1, \bm{P}_2 \right)$ is given by
\begin{align}
\mathcal{I}\left( \bm{P}_1, \bm{P}_2 \right) =\frac{1}{\pi}
\int_0^{T} \frac{ g^{-\left(n_R -1\right)} \sin^{2n_R}\theta}{g_0 +
\sin^2\theta } d\theta, \label{eq:high_mmse:integral}
\end{align}
\noindent where $g= g_0 \Tr \left(\bm{P}_1^{-1} \bm{P}_2 \right)$.
The integral, $\mathcal{I}\left( \bm{P}_1, \bm{P}_2 \right)$, can be
solved in closed form by expanding the ratio of $\sin^2 \theta$
terms in \dref{eq:high_mmse:integral} as a polynomial and
integrating term by term to get the final result in
\dref{eq:high_mmse:integral_solution}.
\begin{table*}[!t]
\normalsize
\begin{align}
\begin{split}
\mathcal{I}\left( \bm{P}_1, \bm{P}_2 \right) &=
\left(-1\right)^{n_R} \frac{g_0^{n_R -1}}{\pi g^{n_R-1}} \left\{
\sqrt{\frac{g_0}{ 1+g_0}} \tan^{-1} \left(
\sqrt{\frac{1+g_0}{g_0}} \tan T \right)  -  \sum_{i=0}^{n_R-1} (-1)^{i} \frac{1}{g_0^i} \right. \\
 &\times \left. \left[\frac{T}{2^{2i}} \left(\frac{2i}{i}\right) + \frac{(-1)^{i} }{2^{2i-1}}  \sum_{k=0}^{i-1} (-1)^{k} \left(\frac{2i}{k}\right) \frac{\sin
\left( 2\left(i - k\right)T \right) }{2\left(i - k \right )} \right]
\right \}. \label{eq:high_mmse:integral_solution}
\end{split}
\end{align}
\par
\hrulefill \vspace{0pt}\\
\end{table*}
We note that, as expected, the diversity order of $n_R-1$ is
observed in both receiver types
and the difference only appears in the array gains.\\
The approximate, high SNR result for ZF in \dref{eq:high_zf:ser1} is
particularly useful since it is simpler than the MMSE version in
\dref{eq:high_mmse:ser1}, and at high SNR the performance of the two
schemes is similar anyway. Hence, \dref{eq:high_zf:ser1} acts as a
useful approximation for both ZF and MMSE and provides a remarkably
compact relationship between SER and the link powers, via the single
function, $\vartheta \left (\bm{P}_1, \bm{P}_2 \right)$.

\section{High SNR Analysis}\label{sec:high_snr_analysis_more_accurate}
In this section, we derive exact high SNR results for the SER of
MMSE and ZF receivers. The work in this section does not employ the
Laplace type approximation used in \dref{eq:high_zf:cf0} and
\dref{eq:high_mmse:cf0} and hence produces exact asymptotics at the
expense of increased complexity. The mathematical details are given
in brief to avoid unnecessary detail.
\subsection{ZF Analysis} \label{subsec:high_snr_ZF_more_accurate}
The conditional cf in \dref{eq:zf:cf3} is a ratio of quadratic forms
in $\bm{h}_2$. As the SNR grows, $\sigma^2\rightarrow 0$ and keeping
only the dominant power of $\sigma^2$ in \dref{eq:zf:cf3} gives
\begin{align}
\phi_{\tilde{Z}}(t | \bm{h}_2) \approx \frac{\bm{h}^H_2 \bm{h}_2}{|
\bm{P}_1 | \left ( \bm{h}^H_2 \bm{P}_1^{-1} \bm{h}_2 \right ) }
\left( \frac{\sigma^2}{-jt} \right)^{n_R-1}.
\label{eq:zf:cf2_more_accu}
\end{align}
Hence, the unconditional cf, when $\sigma^2\rightarrow 0$, becomes
\begin{align}
\phi_{\tilde{Z}}(t) = \tilde{K}_0 \left( \frac{\sigma^2}{-jt}
\right)^{n_R-1} + o \left(\sigma^{2\left(n_R-1\right)}\right),
\label{eq:zf:cf3_more_accu}
\end{align}
\noindent where $o\left(.\right)$ is the standard "little-o"
notation and represents the fact that only the dominant power of
$\sigma^2$ is used in the approximation and
\begin{align}
\tilde{K}_0 = \frac{1}{| \bm{P}_1 |} E \left \{ \frac{\bm{h}^H_2
\bm{h}_2}{\left. \bm{h}^H_2 \bm{P}_1^{-1} \bm{h}_2 \right.} \right
\}. \label{eq:zf:K0_more_accu}
\end{align}
Following the same mgf based procedure to obtain the SER as in Sec.
\ref{sec:high_snr_analysis}, we arrive at the following expression
\begin{align}
\tilde{P}_s = \left(\tilde{G}_{ea}
\bar{\gamma}\right)^{-\tilde{G}_{ed}} +
o\left(\bar{\gamma}^{-\tilde{G}_{ed}} \right),
\label{eq:high_zf:ser1_more_accu}
\end{align}
where, the diversity and array gains are given by
$$
\tilde{G}_{ed} = n_R - 1, \hspace{10mm} \tilde{G}_{ea} = \left(
\tilde{K}_0 \tilde{\mathcal{I}} \right)^{-1/(n_R-1)}. \nonumber
$$
The constant, $\tilde{K}_0$, can be found using Lemma 1 to obtain
the final expression as in \dref{eq:high_zf:K0_more_accu_solution}
\begin{table*}[!t]
\normalsize
\begin{align}
\begin{split}
\tilde{K}_0 = \sum_{i=1}^{n_R} \Upsilon_i P_{i2} + \sum_{1\leq u
\neq v \leq n_R }^{n_R} \left( \Upsilon_u P_{v1} + \Upsilon_v P_{u1}
\right) \left( \frac{\ln \left( \frac{P_{u1}}{P_{u2}} \right) - \ln
\left( \frac{P_{v1}}{P_{v2}} \right)}{\frac{P_{u1}}{P_{u2}}
 -  \frac{P_{v1}}{P_{v2}} } \right)
\label{eq:high_zf:K0_more_accu_solution}
\end{split}
\end{align}
\hrulefill \vspace{0pt}\\
\end{table*}
where
\begin{align}
\Upsilon_i &= \frac{P_{i2}^{n_R-2}}{\prod_{k \neq i }^{n_R}
P_{k1}P_{i2} - P_{i1} P_{k2}}. \label{eq:high_zf:upsilon}
\end{align}
\subsection{MMSE Analysis} \label{subsec:high_snr_MMSE_more_accurate}
Consider the expectation of the conditional cf expression in
\dref{eq:mmse:cf3}. As the SNR grows, $\sigma^2\rightarrow 0$ and
keeping only the dominant power of $\sigma^2$ in \dref{eq:mmse:cf3}
gives
\begin{align}
\phi_{Z}(t) = K_0\left( jt \right) \left( \frac{\sigma^2}{-jt}
\right)^{n_R-1} + o \left(\sigma^{2\left(n_R-1\right)}\right),
\label{eq:high_mmse:cf4_more_accu}
\end{align}
where
\begin{equation}
K_0 \left( s \right) = \frac{1}{|\bm{P}_1|} E \left \{
\frac{\bm{h}_2^H \bm{h}_2 }{ \left. \bm{h}_2^H \bm{P}_1^{-1}
\bm{h}_2  - s \right.} \right \}. \label{eq:high_mmse:cf4}
\end{equation}
Following the mgf based procedure in Sec.
\ref{sec:high_snr_analysis}, the SER at high SNR becomes
\begin{align}
P_s \approx \frac{1}{\pi} \int_0^{T} \left( \frac{\sigma^2 \sin^2
\theta}{g}\right)^{n_R-1} K_0 \left(-\frac{g}{\sin^2 \theta} \right)
d\theta. \label{eq:high_mmse:ser0_more_accu}
\end{align}
From \dref{eq:high_mmse:ser0_more_accu}, the approximate SER can be
written in terms of the diversity gain and array gain as
\begin{align}
P_s = \left(G_{ea} \bar{\gamma}\right)^{-G_{ed}} +
o\left(\bar{\gamma}^{-G_{ed}} \right),
\label{eq:high_mmse:ser1_exact}
\end{align}
where
$$
G_{ed}=n_R - 1, \hspace{10mm} G_{ea} = \left(\mathcal{I}_e \left(
\bm{P}_1, \bm{P}_2 \right)\right)^{-1/(n_R-1)}, \nonumber
$$
\begin{align}
\mathcal{I}_e \left( \bm{P}_1, \bm{P}_2 \right) \!=\! \frac{1}{\pi}
\int_0^{T} \!\!\! \left(\! \frac{ \sin^2 \theta}{g}
\!\right)^{n_R-1} \!\!\! \!K_0 \left(- \frac{g}{\sin^2 \theta}
\right) d\theta. \label{eq:high_mmse:integral_more_accu}
\end{align}

Using Lemma 1, $K_0 \left( -s \right)$ can be given by
\begin{equation}
K_0 \left( -s \right) = - \int_0^\infty \frac{\partial}{\partial
\theta_1 } \left[\frac{e^{-s \theta_2}}{| \bm{P}_1 + \theta_1
\bm{P}_2\bm{P}_1 + \theta_2 \bm{P}_2 |} \right]_{\theta_1=0}
d\theta_2, \label{eq:high_mmse:ratio_expectation}
\end{equation}
which can be simplified to obtain
\begin{align}
\begin{split}
K_0 \left( -s \right) = \int_0^\infty e^{-s \theta_2} &\left (
\sum_{i=1}^{n_R} \frac{P_{i2}\Upsilon_i}{P_{i2}\theta_2 +
P_{i1}}\right)  \left ( \sum_{i=1}^{n_R}
\frac{P_{i1}P_{i2}}{P_{i2}\theta_2 + P_{i1}}\right) d\theta_2,
\end{split}
\label{eq:high_mmse:ratio_expectation2}
\end{align}
where $\Upsilon_i$ is given in \dref{eq:high_zf:upsilon}. Equation
\dref{eq:high_mmse:ratio_expectation2} can be solved in closed form
to give
\begin{align}
\begin{split}
K_0 \left( -s \right) \!=\! \sum_{i=1}^{n_R} \left \{ e^{s
\frac{P_{i1}}{P_{i2}}} E_1 \left(s \frac{P_{i1}}{P_{i2}} \right)\!\!
\left( \Phi_i - s \Upsilon_i P_{i1}\right) + \Upsilon_i P_{i2}
\right \},
\end{split}
\label{eq:high_mmse:ratio_expectation3}
\end{align}
where
\begin{align}
\begin{split}
\Phi_i &= \sum_{k \neq i}^{n_R} \frac{P_{i2}\Upsilon_i P_{k1}P_{k2}
+ P_{k2}\Upsilon_k P_{i1}P_{i2}}{P_{k1}P_{i2} - P_{k2}P_{i1}}.
\end{split}
\label{eq:high_mmse:psi_more_accu}
\end{align}
Substituting $s=g/\sin^2\theta$ in
\dref{eq:high_mmse:ratio_expectation3} and then substituting $K_0
\left( -\frac{g}{\sin^2 \theta} \right)$ in
\dref{eq:high_mmse:integral_more_accu} and integrating over $\theta$
gives the result in \dref{eq:high_mmse:integral_more_accu_solution}
\begin{table*}[!t]
\normalsize
\begin{align}
\begin{split}
\mathcal{I}_e \left( \bm{P}_1, \bm{P}_2 \right) =
\tilde{\mathcal{I}} \left ( \sum_{i=1}^{n_R} \Upsilon_i P_{i2}
\right)+ \frac{1}{\pi g^{\left. n_R-1 \right.}}  \sum_{i=1}^{n_R}
\left\{ \Phi_i H \left(n_R -1 , g \frac{P_{i1}}{P_{i2}}\right) -
\Upsilon_i P_{i1} H \left(n_R -2 , g \frac{P_{i1}}{P_{i2}}\right)
\right \}
\end{split}
\label{eq:high_mmse:integral_more_accu_solution}
\end{align}
\par
\hrulefill \vspace{0pt}\\
\end{table*}
where
\begin{align}
\begin{split}
H \left(m , a \right) = \int_0^{T} e^{\frac{a}{\sin^2\theta}} E_1
\left( \frac{a}{\sin^2\theta} \right) \left. \sin^{2m}\theta \right.
d\theta.
\end{split}
\label{eq:high_mmse:integral_solution_more_accu_integral}
\end{align}
Clearly the exact asymptotics, especially for the MMSE case, are
substantially more complex than the approximations in Sec.
\ref{sec:high_snr_analysis}. Also, the relationship between SER and
the link powers is far more involved.
\section{Numerical Results}\label{sec:numerical_results}
\begin{figure}[h]
\centerline{\includegraphics*[scale=0.65]{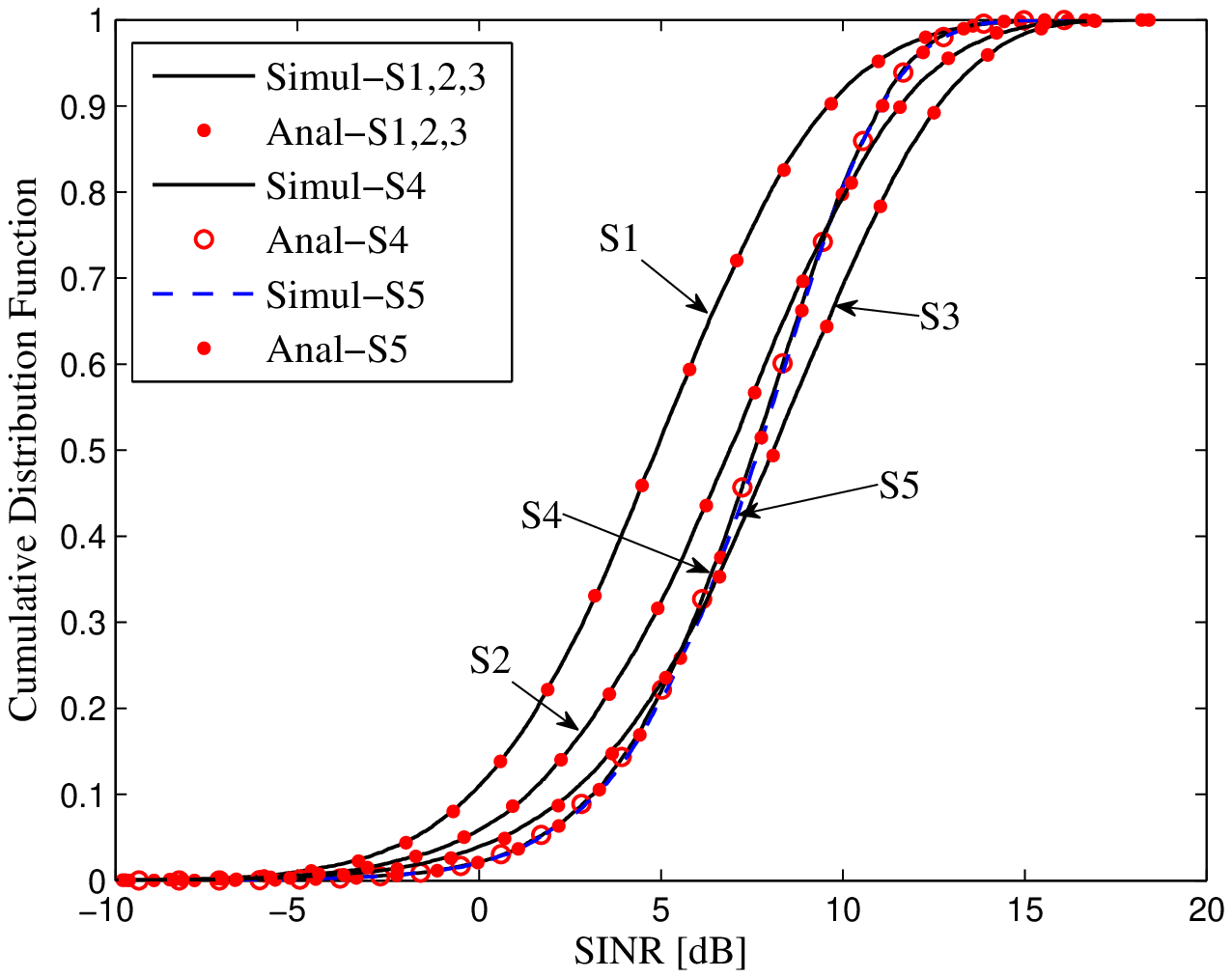}}
\caption{Analytical and simulated cdfs for the output SINR of an
MMSE receiver for scenarios S1-S5 listed in Table 1 at $\rho=5$ dB
with $\varsigma=1$.} \label{fig:cdfs_s1-s5_mmse}
\end{figure}
\begin{figure}[h]
\centerline{\includegraphics*[scale=0.65]{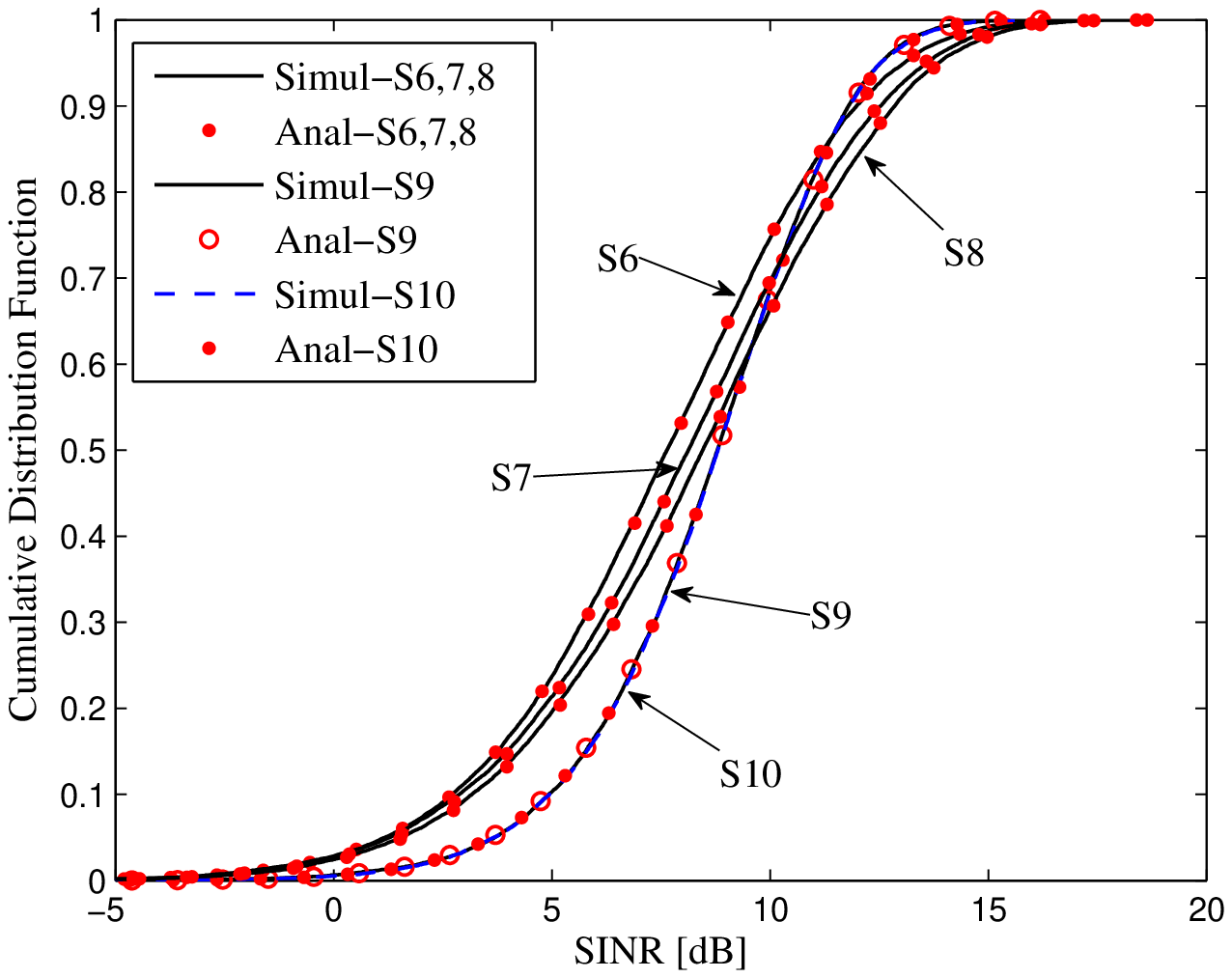}}
\caption{Analytical and simulated cdfs for the output SINR of an
MMSE receiver for scenarios S6-S10 listed in Table 1 at $\rho=5$ dB
with $\varsigma=20$.} \label{fig:cdfs_s6-s10_mmse}
\end{figure}

\begin{figure}[h]
\centerline{\includegraphics*[scale=0.65]{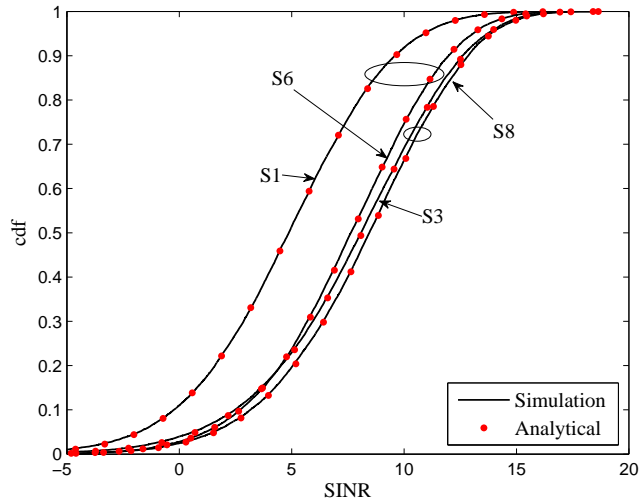}}
\caption{Analytical and simulated cdfs for the output SINR of an
MMSE receiver for scenarios (S1,S6) and (S3,S8) in Table 1.}
\label{fig:cdfs_s1s6_s3s8_mmse}
\end{figure}
\begin{table}[Parameters for Figures]
    \caption{Parameters for Figures \ref{fig:cdfs_s1-s5_mmse} and \ref{fig:cdfs_s6-s10_mmse}}
    \centering
    \begin{tabular}{ c | l | l | c }
    \hline \hline
    & \multicolumn{2}{|c|}{Decay Parameter} & \\ \cline{2-3}
    Sc. No.  & Desired & Interfering &  $\varsigma$  \\ \hline
    S1 &  $\alpha=0.2$ & $\alpha=0.2$ & 1\\
    S2 &  $\alpha=0.2$ & $\alpha=1$ & 1 \\
    S3 &  $\alpha=0.2$ & $\alpha=5$ & 1 \\
    S4 &  $\alpha=1$ & $\alpha=1$ & 1 \\
    S5 &  $\alpha=1$ & $\alpha=0.2$ & 1 \\ \hline
    S6 & $\alpha=0.2$ & $\alpha=0.2$ & 20  \\
    S7  & $\alpha=0.2$ & $\alpha=1$ & 20\\
    S8  & $\alpha=0.2$ & $\alpha=5$ & 20\\
    S9  & $\alpha=1$ & $\alpha=1$ & 20\\
    S10 & $\alpha=1$ & $\alpha=0.2$ & 20\\
    \hline
    \end{tabular}
\label{table:mmse_zf_scenarios} 
\end{table}
\begin{figure}[h]
\centerline{\includegraphics*[scale=0.65]{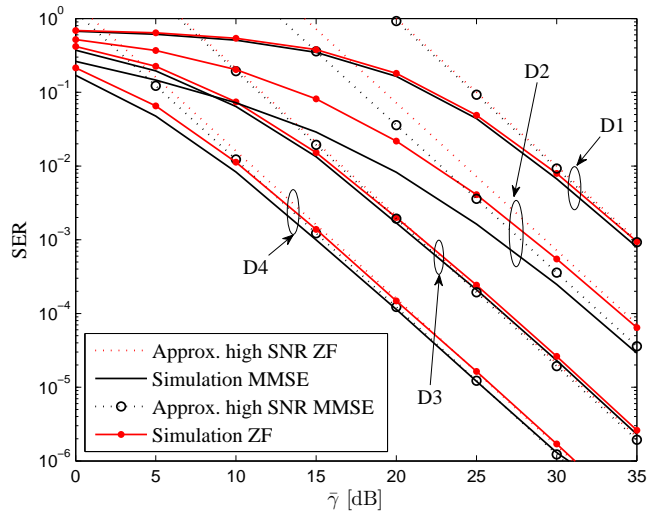}}
\caption{SER and high SNR approximations for MMSE/ZF receivers using
QPSK modulation in flat Rayleigh fading for four arbitrary drops.}
\label{fig:ser_mmse_zf_laplace_high_snr}
\end{figure}

\begin{figure}[h]
\centerline{\includegraphics*[scale=0.65]{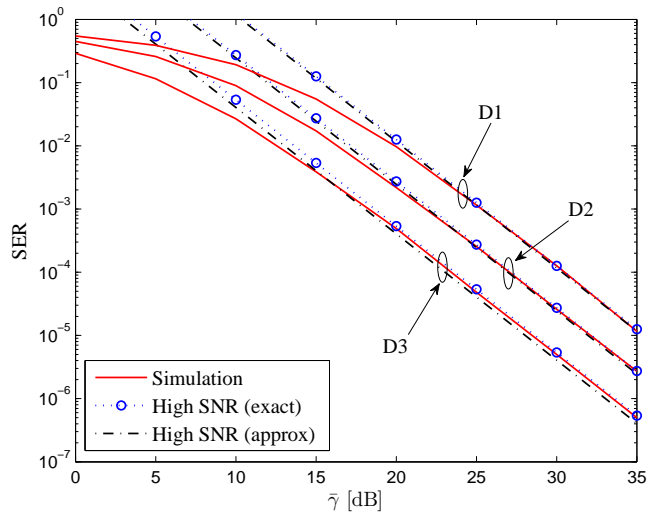}}
\caption{SER of a ZF receiver using QPSK modulation in Rayleigh flat
fading for three arbitrary drops.} \label{fig:ser_zf}
\end{figure}

\begin{figure}[h]
\centerline{\includegraphics*[scale=0.65]{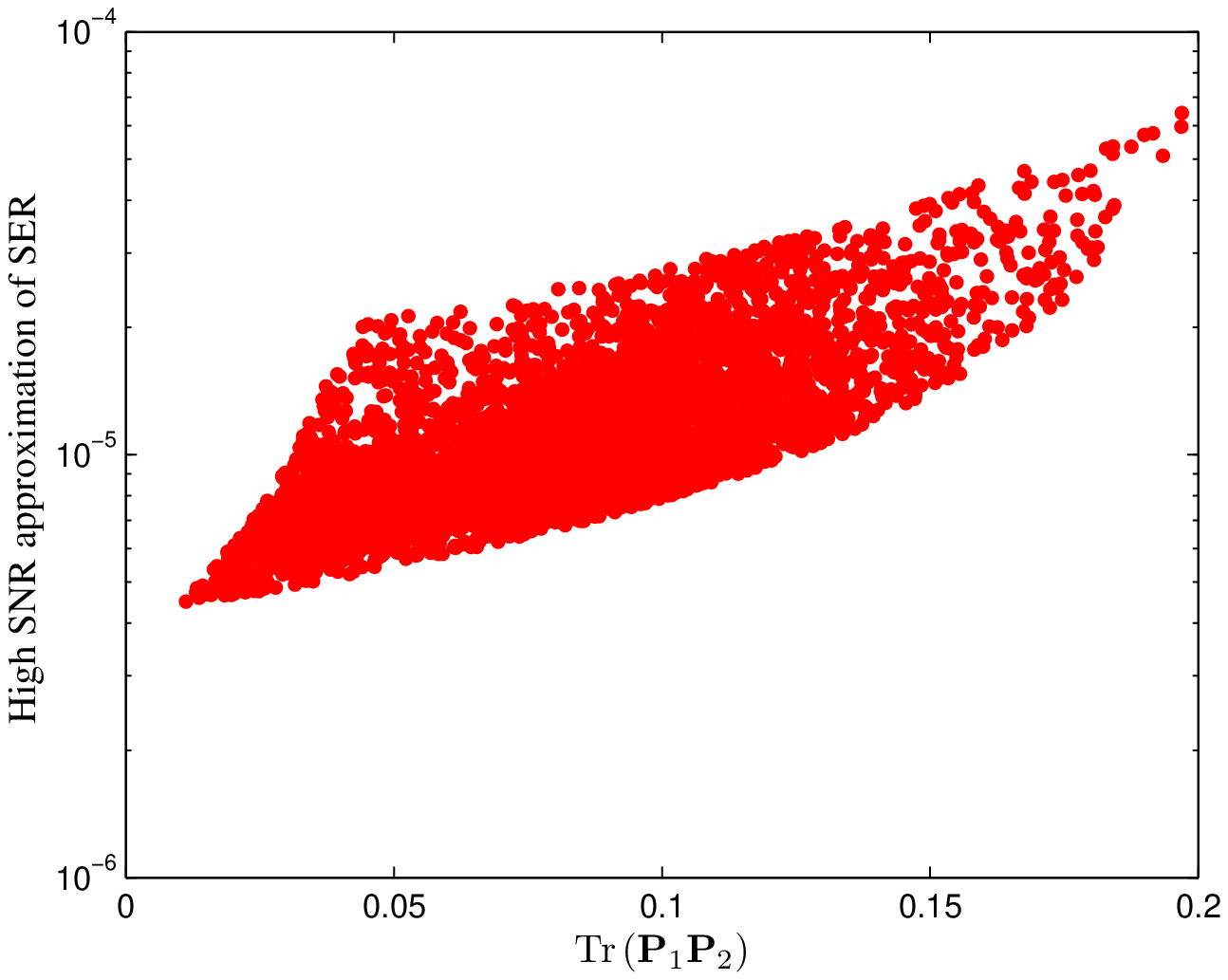}}
\caption{SER of a MMSE receiver for QPSK modulation in flat Rayleigh
fading at $\bar{\gamma}=35$ dB.} \label{fig:mmse:cloud}
\end{figure}
In this section, we verify the analysis by Monte Carlo simulations
using the network MIMO scenario in Fig. \ref{sec1:fig:macro}
\cite{Dush11}. We also consider some special cases of $\bm{P}_1$ and
$\bm{P}_2$ in order to investigate the effect of the macrodiversity
powers on performance. For the two-user system in Fig.
\ref{sec1:fig:macro}, we consider the desired user to be user 1 and
parameterize the system by three parameters. The average received
signal to noise ratio is defined by $\rho= \left. \Tr\left(
\bm{P}_1\right) \right. / n_R \sigma^2$. The total signal to
interference ratio is defined by $\varsigma = \Tr\left(
\bm{P}_1\right) / \Tr\left( \bm{P}_2\right)$. The spread of the
signal power across the three antennas is assumed to follow an
exponential profile, as in \cite{Gao98}, so that a range of
possibilities can be covered with only one parameter. The
exponential profile is defined by
\begin{align}
P_{ik} &= K_k \left( \alpha \right) \alpha^{i-1},
\label{eq:mmse_zf_dual:numerical_results1}
\end{align}
for receive antenna i, source k where
\begin{align}
K_k \left( \alpha \right) &= \Tr\left( \bm{P}_k\right) /
\left(1+\alpha + \alpha^2 \right), \quad k=1,2,
\end{align}
and $\alpha > 0$ is the parameter controlling the uniformity of the
powers across the antennas. Note that as $\alpha \rightarrow 0$ the
received power is dominant at the first antenna, as $\alpha$ becomes
large $\left( \alpha \gg 1 \right)$ the third antenna is dominant
and as $\alpha \rightarrow 1$ there is an even spread, as in the
standard microdiversity scenario. Although we consider
microdiversity (S4 and S9), this is in the context of exploring the
effect of different $\bm{P}$ matrix structures. Physically, it is
not sensible to directly compare microdiversity with macrodiversity
as they involve different system structures. In microdiversity,
there may be multiple users communicating with a single array at a
single BS. In macrodiversity, the users may be communicating with
distributed antennas located at different BS sites which are
back-hauled together to enable joint transmission/reception. In
Figs. \ref{fig:cdfs_s1-s5_mmse}-\ref{fig:cdfs_s6-s10_mmse} we show
cdf results for the ten scenarios given in Table
\ref{table:mmse_zf_scenarios}. In Fig. \ref{fig:cdfs_s1-s5_mmse}, we
see that S1 has the worst cdf since the sharply decaying power
profile is identical for both desired and interfering source. Hence,
there is reduced diversity, as most of the signal strength is seen
at one antenna, and there is strong interference at each antenna.
Scenario S3 is best at high SINR since in this interference limited
situation $\left(\varsigma=1\right)$ it is best to have at least one
antenna where there is minimal interference. This occurs with S3 as
the power profiles are opposing and the strongest desired signal
aligns with the weakest interferer. At low SINR, scenarios S4 and S5
are slightly better than S3 as they have full diversity (equal power
at each antenna) which is beneficial in this SINR region. In Fig.
\ref{fig:cdfs_s6-s10_mmse}, similar results are observed with
diversity being important at lower SINR (where S9 and S10 are the
best) and interference reduction being important at high SINR (where
S8 is best). In Fig. \ref{fig:cdfs_s1s6_s3s8_mmse}, we consider the
effect of $\varsigma$ on two different macrodiversity scenarios in
Table \ref{table:mmse_zf_scenarios}. In particular, we have shown
results for S1 and S6 and S3 and S8. As expected, S1 and S3 have
lower SINRs than S6 and S9 due to increased interference. However,
S1 is far more sensitive to $\varsigma$ than S3. This is because S3
has opposing power profiles for the desired and interfering users so
that the two sources are more orthogonal and
interference plays a smaller part in performance.\\
Next, we consider the high SNR results in Secs.
\ref{sec:high_snr_analysis} and
\ref{sec:high_snr_analysis_more_accurate}. In Fig.
\ref{fig:ser_mmse_zf_laplace_high_snr}, the MMSE/ZF receivers are
considered for four drops (D1, D2, D3 and D4) of two users in the
shaded coverage area of Fig. \ref{sec1:fig:macro}. Each user is
dropped at a different random location (uniformly generated over the
coverage area) and random lognormal shadow fading and path loss is
considered where $\sigma_{SF}=8$dB (standard deviation of shadow
fading) and $\gamma=3.5$ (path loss exponent). Hence, each user has
a different distance and shadow fade to each BS and each drop
results in a new $\bm{P}$ matrix. The transmit power of the sources
is scaled so that all locations in the coverage area have a maximum
received SNR greater than $3$dB, at least $95\%$ of the time. The
maximum SNR is taken over the 3 BSs. For all four drops, the high
SNR approximations from Sec. \ref{sec:high_snr_analysis} are shown
to be very accurate for SERs below $10^{-2}$, although for drop D2
the results are less tight. Note that drop D2 has the greatest
difference between the MMSE and ZF results. In general, the gap
between MMSE and ZF results can be assessed by a comparison of
\dref{eq:high_zf:integral} with \dref{eq:high_mmse:integral}. Here,
it can be seen that the approximate asymptotics are the same for ZF
and MMSE when $g_0=0$. Hence, scenarios where $g_0$ is large, i.e.,
$\Tr \left( \bm{P}_1^{-1}\bm{P}_2 \right)\approx 0$, will create
substantial differences between the two receivers. In Fig.
\ref{fig:ser_mmse_zf_laplace_high_snr}, drop D2 had the smallest
value of $\Tr \left( \bm{P}_1^{-1}\bm{P}_2 \right)$ and hence showed
the greatest difference. Note that for $\Tr \left(
\bm{P}_1^{-1}\bm{P}_2 \right)$ to be small, $P_{i1} \gg P_{i2}$ is
required for $i=1,2, \dots,n_R$. Hence, the power profiles for users
1 and 2 must be ``parallel'' in some sense, with any large value of
$P_{i2}$ aligning with an even larger value of $P_{i1}$. In these
``parallel'' scenarios, MMSE and ZF results can exhibit greater
differences.\\
This methodology is used again in Fig. \ref{fig:ser_zf} for three
drops and ZF results are shown. Again, the asymptotic results show
good agreement at SERs below $10^{-2}$. Furthermore, the difference
between the approximations in Sec. \ref{sec:high_snr_analysis} and
the exact asymptotics in Sec.
\ref{sec:high_snr_analysis_more_accurate} is shown to be minor.
Hence the simple SER forms in  \dref{eq:high_zf:ser1} and
\dref{eq:high_mmse:ser1} are particularly useful. Note that the
power matrices, $\bm{P}_1$ and $\bm{P}_2$, are completely general,
with the sole constraint being $P_{ik} \geq 0$, $\forall i,k$. As a
result, it is likely that some combinations of powers can be found
that will cause the approximate SERs to lose accuracy. However, in
all the scenarios considered and all random drops simulated (see
also \cite{Dush11}) the approximations have shown similar accuracy
to the results in Figs. \ref{fig:ser_mmse_zf_laplace_high_snr} and
\ref{fig:ser_zf}.\\
In multiuser systems it is well-known that two users may be
successfully detected if their channels are approximately
orthogonal. In the context of a dual user system, where only the
channel powers are considered, the analog would be that $\Tr \left(
\bm{P}_1 \bm{P}_2 \right)$ is small. To investigate this
relationship, we generate a large number of random power matrices
with a fixed total power. For user 1, the powers are generated by
\dref{eq:mmse_zf_dual:numerical_results1} with $\alpha=0.2$. For
user 2, the powers are independent uniform random variables which
are scaled so that $\varsigma=1$. For each pair,
$\left(\bm{P}_1,\bm{P}_2 \right)$, we compute $\Tr \left( \bm{P}_1
\bm{P}_2 \right)$ and the approximate SER of user 1 using using
\dref{eq:high_mmse:ser1}. The results are plotted in Fig.
\ref{fig:mmse:cloud}. As expected, SER increases with $\Tr \left(
\bm{P}_1 \bm{P}_2 \right)$, although there is wide variation in the
band of SER results. In comparison, the new metric, $\vartheta \left
(\bm{P}_1, \bm{P}_2 \right)$, has a one-to-one relationship with the
approximate SER and carries far more information than ad-hoc
measures such as $\Tr \left( \bm{P}_1 \bm{P}_2 \right)$.
%
%
%
%

%
\section{Conclusion}\label{sec:conclusion}
In this paper, we derived the exact cdf of the output SINR/SNR for
MMSE and ZF receivers in the presence of a single interfering user
with an arbitrary number of receive antennas. To the best of our
knowledge, this represents the first exact analysis of linear
combining in macrodiversity systems. Although not shown for reasons
of space, the slightly simpler problem of obtaining the associated
pdf can also be handled since the pdf expression in
\dref{eq:zf:pdf1} is inherently computed en-route to the cdf in
\dref{eq:zf:cdf1}. Numerical examples demonstrate the validity of
the analysis across arbitrary drops and channels. This suggests that
the analysis is also numerically robust. A high SNR analysis reveals
simple SER results for both MMSE and ZF which provide insights into
both diversity and array gain. It also provides a functional link
between the performance of the macrodiversity system and the link
SNRs.
\appendices
\section{Background Results}\label{app:A00}
The following lemma gives a compact method to calculate the expected
value of a ratio of random variables with arbitrary integer powers
\cite{Suwa72}.
\vspace*{3mm}
\begin{Lemma}
Let $U_1$, $U_2$ and $Z$  be three continuous random variables such
that $P(U_2
> 0)=1$, $Z \geq 0$ and $Z=\frac{U_1^m}{U_2^n}$. Assuming that there exists a joint moment generating
function (mgf) for $U_1$ and $U_2$, denoted
$\mathcal{M}(\theta_1,\theta_2)=E\left( e^{\theta_1 U_1 + \theta_2
U_2}\right)$, then, for all positive integer values of $m$ and $n$,
the expected value of $Z$ is given by\vspace*{0mm}
\begin{equation}
E \left \{ \frac{U_1^m}{U_2^n} \right \} = \frac{1}{(n-1)!}
\int_0^\infty  z^{n-1}\left. \frac{\partial^m \mathcal{M}(\theta,
-z)}{\partial \theta^m}\right|_{\theta=0} dz.  \nonumber
\end{equation}
\vspace*{0mm} \label{lemma:expected_value_of_ratios}
\end{Lemma}
The following n-dimensional complex Gaussian integral identity will
be used extensively.
\begin{Lemma}
Let $\bm{A}$ be an arbitrary $n \times n$ complex Hermitian positive
definite matrix. Then, the following integral identity holds.
\vspace{0mm}
\begin{equation}
\int_{-\infty}^{\infty} \ldots \int_{-\infty}^{\infty} e^{-\bm{x}^H
\bm{A} \normalsize{\bm{x}}} dx_1 \ldots dx_n = \frac{\pi^n}{\left|
\bm{A} \right|}, \nonumber
\end{equation}
where the complex $n \times 1$ vector $\bm{x}=[x_1, \ldots, x_n]^T$,
$dx_i=dx_{iI}dx_{iQ}$, $x_{iI}=\textup{Re}(x_i)$ and
$x_{iQ}=\textup{Im}(x_i)$. \label{lemma:complex_gaussian}
\end{Lemma}
Finally, we give the following partial fraction expansion from
elementary algebra,\vspace{0mm}
\begin{equation}
\frac{1}{\prod_{i=1}^{n_R} \left ( a_i - jtb_i \right )} =
\sum_{i=1}^{n_R} \frac{A_i}{\left . \frac{a_i}{b_i} - jt \right. },
\label{eq:partial_fraction}
\end{equation}
where $j^2 = -1$ and
\begin{equation}
A_i = \frac{b_i^{n_R-2}}{\prod_{k \neq i}^{n_R} \left ( b_i a_k -
a_i b_k \right )}.
\end{equation}
\section{ZF Analysis}\label{app:A0}
From \cite{Proakis01},  the full cf can be obtained by averaging the
conditional cf in \dref{eq:zf:cf3} over $\bm{h}_2$. Using Lemma
\ref{lemma:expected_value_of_ratios},  the full cf is given by
\begin{equation}
\phi_{\tilde{Z}}(t) = - \frac{1}{\left|\bm{D}\right|} \int_0^\infty
\left. \frac{\partial E \left \{ e^{-\theta_1 z_1 - \theta_2 z_2}
\right \} }{\partial \theta_1}\right|_{\theta_1=0} d\theta_2,
\label{eq:zf:cf3:1}
\end{equation}
\noindent where $z_1=\bm{h}_2^H \bm{h}_2$ and $z_2= \bm{h}_2^H
\bm{D}^{-1} \bm{h}_2$. Note that the dummy variables are $\theta_1,
\theta_2 \geq 0$ and we have used a slight variation of Lemma
\ref{lemma:expected_value_of_ratios}, in which the joint mgf has
negative coefficients, for convenience. Using the pdf of $\bm{h}_2$
in \dref{eq:pdfs} to evaluate the expectation in \dref{eq:zf:cf3:1}
and using the Gaussian integral identity in Lemma
\ref{lemma:complex_gaussian} and a few simplifications, we obtain
the following result.
\begin{align}
\phi_{\tilde{Z}}(t) = - \int_0^\infty \frac{\partial}{\partial
\theta_1 } \left[\frac{1}{\left| \bm{D}_1 - jt \bm{D}_2 \right|}
\right]_{\theta_1=0} d\theta_2,
 \label{eq:zf:cf4}
\end{align}
\noindent where $\bm{D}_1= \bm{I} + \theta_1 \bm{P}_2 + \theta_2
\bm{P}_2$ and $\bm{D}_2= \frac{1}{\sigma^2} \bm{P}_1 \left(\bm{I} +
\theta_1 \bm{P}_2 \right)$. From \cite{Proakis01}, the pdf and cdf
of $\tilde{Z}$ are
\begin{equation}
f_{\tilde{Z}}(z) = \frac{1}{2\pi}\int_{-\infty}^\infty
\phi_{\tilde{Z}}(t) e^{-jtz} dt, \label{eq:zf:pdf1}
\end{equation}
and
\begin{equation}
F_{\tilde{Z}}(z) = \frac{1}{2\pi} \int_0^z \int_{-\infty}^\infty
\phi_{\tilde{Z}}(t) e^{-jtx} dt dx. \label{eq:zf:cdf1}
\end{equation}\\
By substituting \dref{eq:zf:cf4} into \dref{eq:zf:pdf1} and
\dref{eq:zf:cdf1} multiple integral forms for the pdf and cdf of
$\tilde{Z}$ are obtained as
\begin{align}
f_{\tilde{Z}}(z) = -\frac{1}{2\pi} \int_{-\infty}^\infty
\int_0^\infty \frac{\partial}{\partial \theta_1 } \left[
\frac{e^{-jtz}}{\left| \bm{D}_1 - j t \bm{D}_2 \right|}
\right]_{\theta_1=0} d\theta_2 dt, \label{eq:zf:pdf2}
\end{align}
\begin{align}
F_{\tilde{Z}}(z) = -\frac{1}{2\pi} \int_0^z \int_{-\infty}^\infty
\int_0^\infty \frac{\partial}{\partial \theta_1 } \left[
\frac{e^{-jtx}}{\left| \bm{D}_1 - j t \bm{D}_2 \right|}
\right]_{\theta_1=0} d\theta_2 dt dx. \label{eq:zf:cdf2}
\end{align}
Since $\bm{D}_1$ and $\bm{D}_2$ are diagonal, we can further
simplify the expression in \dref{eq:zf:cdf2} with the substitutions,
$a_i= 1 + \theta_1 P_{i2} + \theta_2 P_{i2}$ and
$b_i=\frac{1}{\sigma^2} P_{i1} \left( 1 + \theta_1 P_{i2} \right )$.
Then, the integrand in \dref{eq:zf:cdf2}, before differentiation,
can be written as $\tilde{J}_0 =\frac{e^{ - jtx}} {\prod_{i=1}^{n_R}
\left ( a_i - jt b_i \right )}$. Hence,
\vspace{0mm}
\begin{align}
\begin{split}
F_{\tilde{Z}}(z) = -\frac{1}{2\pi} \int_0^z  \int_{-\infty}^\infty
\int_0^\infty  \left. \frac{\partial \tilde{J}_0}{\partial \theta_1
} \right |_{\theta_1=0} d\theta_2 dt dx. \label{eq:zf:cdf3}
\end{split}
\end{align}
\noindent Since the limits of integration are independent of
$\theta_1$, we can interchange the order of differentiation and
first perform the integration over $t$. To obtain this integral, we
use the partial fraction expansion of $\tilde{J}_0$ from
\dref{eq:partial_fraction} and apply the following integral identity
from \cite{GradRzy00},
\vspace{0mm}
\begin{eqnarray}
\int_{-\infty}^{\infty} \frac{e^{-jpx}}{\left (\beta - j x \right
)^v} dx = \begin{cases} \frac{2\pi p^{v-1}e^{-\beta p}}{\Gamma \left
(v \right )}& \hspace{0mm}  p > 0  \nonumber \\
0& \hspace{0mm}  p < 0,
\end{cases} \hspace{10mm}
\hspace{0mm} \left[ Re(v) > 0, Re(\beta)> 0 \right].
\label{eq:comlex_integral}
\end{eqnarray}
This gives the result,
\begin{align}
\begin{split}
F_{\tilde{Z}}(z) = -\int_0^z \int_0^\infty \left. \frac{\partial
\tilde{J}_1}{\partial \theta_1 } \right|_{\theta_1=0} d\theta_2 dx,
\end{split}
\label{eq:zf:cdf33}
\end{align}
\noindent where $\tilde{J}_1=\sum_{i=1}^{n_R} A_i e^{-
\frac{a_i}{b_i}x}$. Differentiating $\tilde{J}_1$ term by term and
setting $\theta_1=0$ gives \dref{eq:zf:J1}
%
%
\begin{table*}\small
\begin{align}
\left. \frac{\partial \tilde{J}_1}{\partial \theta_1 }
\right|_{\theta_1=0} = \sigma^2 \sum_{i=1}^{n_R}
e^{-\tilde{\alpha}_i x - \tilde{\beta}_i x \theta_2 } \left\{
\frac{P_{i1}^{n_R-2} \left(  x \theta_2 \sigma^2 P_{i2}^2
P_{i1}^{-1} + (n_R-2)P_{i2} \right) }{\prod_{k\neq i}^{n_R} \left(
\tilde{n}_{ik} + \theta_2 \tilde{m}_{ik} \right) } -
\frac{P_{i1}^{n_R-2}}{\prod_{k\neq i}^{n_R} \tilde{n}_{ik} +
\theta_2 \tilde{m}_{ik}} \left[ \sum_{k\neq i}^{n_R}
\frac{\tilde{\gamma}_{ik} + \theta_2
\tilde{\delta}_{ik}}{\tilde{n}_{ik} + \theta_2 \tilde{m}_{ik}}
\right] \right \} \label{eq:zf:J1}
\end{align}
\hrulefill
\end{table*}
where $\tilde{\gamma}_{ik} = \left( P_{i1} - P_{k1} \right) \left(
P_{i2}+ P_{k2} \right)$ and $\tilde{\delta}_{ik} = \left( P_{i1} -
P_{k1}\right) P_{i2}P_{k2}$. Also
\begin{subequations}
\begin{align}
\tilde{n}_{ik}&= \left( P_{i1} - P_{k1}\right),  \qquad \tilde{m}_{ik} = \left(P_{i1}P_{k2} - P_{k1}P_{i2}\right),\\
\tilde{\alpha}_i &=  \frac{\sigma^2}{P_{i1}}, \qquad \tilde{\beta}_i
= \frac{\sigma^2 P_{i2}}{P_{i1}}.
\end{align}
\label{eq:zf:basic_constants0}
\end{subequations}
\noindent Using \dref{eq:partial_fraction}, the product in the
denominator of \dref{eq:zf:J1} can be expanded as
\vspace{0mm}
\begin{align}
\frac{1}{\prod_{k\neq i}^{n_R} \left(\tilde{n}_{ik} + \theta_2
\tilde{m}_{ik}\right) } &= \sum_{k\neq i}^{n_R}
\frac{\tilde{\eta}_{ik}}{\left. \tilde{n}_{ik} + \theta_2
\tilde{m}_{ik} \right. }, \label{eq:zf:Bik}
\end{align}
where
\begin{align}
\tilde{\eta}_{ik} = \frac{\tilde{m}_{ik}^{n_R-2}}{\prod_{l\neq
i,k}^{n_R} \left ( \tilde{n}_{il}\tilde{m}_{ik} -
\tilde{n}_{ik}\tilde{m}_{il} \right )}.
\end{align}
Substituting \dref{eq:zf:Bik} in \dref{eq:zf:J1} gives
\dref{eq:zf:J2}
\begin{table*}\small
\begin{align}
\left. \frac{\partial \tilde{J}_1}{\partial \theta_1 }
\right|_{\theta_1=0} = \sigma^2 \sum_{i=1}^{n_R} \sum_{k\neq
i}^{n_R} e^{-\tilde{\alpha}_i x - \tilde{\beta}_i x \theta_2 }
\left\{ \frac{x \theta_2 \sigma^2 P_{i2}^2 P_{i1}^{n_R-3}
\tilde{\eta}_{ik}}{\left ( \tilde{n}_{ik} + \theta_2 \tilde{m}_{ik}
\right )} - \frac{\tilde{\varphi}_{ik}}{\left ( \tilde{n}_{ik} +
\theta_2 \tilde{m}_{ik} \right )} - \frac{P_{i1}^{n_R-2}
\tilde{\eta}_{ik}\tilde{\xi}_{ik}}{\left ( \tilde{n}_{ik} + \theta_2
\tilde{m}_{ik} \right )^2} \right \}. \label{eq:zf:J2}
\end{align}
\end{table*}
where the constants are given by \vspace{0mm}
\begin{align}
\tilde{\Delta}_i &= \sum_{k \neq i}^{n_R} \frac{\left(
P_{i1}P_{i2}P_{k2} - P_{k1}P_{k2}P_{i2}\right)}{\left( P_{i1}P_{k2}
- P_{i2}P_{k1}\right)}, \\
%
%
\tilde{\xi}_{ik} &= \frac{\left( P_{i1} - P_{k1}\right) \left(
P_{i1}P_{k2}^2 - P_{k1}P_{i2}^2 \right)}{\left( P_{i1}P_{k2} -
P_{i2}P_{k1}\right)},\\
\tilde{\zeta}_{ik} &= \tilde{m}_{ik} \sum_{l \neq i,l \neq k}^{n_R}
\frac{ \tilde{\eta}_{ik} \tilde{\xi}_{il} + \tilde{\eta}_{il}
\tilde{\xi}_{ik} }{\left( \tilde{n}_{il} \tilde{m}_{ik} -
\tilde{n}_{ik} \tilde{m}_{il} \right)}.
\end{align}
Substituting $\left. \frac{\partial \tilde{J}_1}{\partial \theta_1 }
\right|_{\theta_1=0}$ from \dref{eq:zf:J2} in to the cdf expression
in \dref{eq:zf:cdf33} gives \dref{eq:zf:cdf4}.
\begin{table*}\small
\begin{align}
F_{\tilde{Z}}(z) = -\sigma^{2}\int_0^z \int_0^\infty
\sum_{i=1}^{n_R} \sum_{k\neq i}^{n_R} e^{ -\tilde{\alpha}_i x -
\tilde{\beta}_i x \theta_2} \left\{ \frac{x \theta_2 \sigma^2
P_{i2}^2 P_{i1}^{n_R-3} \tilde{\eta}_{ik}}{\left ( \tilde{n}_{ik} +
\theta_2 \tilde{m}_{ik} \right )} -
\frac{\tilde{\varphi}_{ik}}{\left ( \tilde{n}_{ik} + \theta_2
\tilde{m}_{ik} \right )} - \frac{P_{i1}^{n_R-2}
\tilde{\eta}_{ik}\tilde{\xi}_{ik}}{\left ( \tilde{n}_{ik} + \theta_2
\tilde{m}_{ik} \right )^2} \right \} d\theta_2 dx.
\label{eq:zf:cdf4}
\end{align}
\hrulefill
\end{table*}
The desired cdf in \dref{eq:zf:cdf4} is rewritten in
\dref{eq:zf:cdf5} where
\begin{subequations}
\begin{align}
\tilde{\varphi}_{ik} &= P_{i1}^{n_R-2} \left (
\tilde{\Delta}_i \tilde{\eta}_{ik} - (n_R-2)P_{i2} \tilde{\eta}_{ik} + \tilde{\zeta}_{ik} \right), \\
\tilde{\psi}_{ik} &= \tilde{\eta}_{ik} \tilde{\xi}_{ik}
P_{i1}^{n_R-2}, \\
\tilde{\omega}_{ik} &= -P_{i1}^{n_R-3} \tilde{\eta}_{ik} P_{i2}^2
\end{align}
\label{eq:zf:three_constants}
\end{subequations}
Note that when $n_R=2$, $\tilde{\zeta}_{ik}=0$ and
$\tilde{\eta}_{ik}=1$ for all $i,k$. The cdf in \dref{eq:zf:cdf5}
contains three types of double integral defined by \vspace{0mm}
\begin{align}
\tilde{I_1}\left(a,b,c,d,x \right) &= \int_0^x \int_0^\infty
\frac{e^{-b t - d t \theta }}{a + c \theta} d\theta dt,
\label{eq:zf:double_integral1}\\
\tilde{I_2}\left(a,b,c,d,x \right) &= \int_0^x \int_0^\infty
\frac{e^{-b t - d t \theta }}{\left( a + c \theta \right )^2}
d\theta dt,\label{eq:zf:double_integral2}\\
%
%
%
\tilde{I_3} \left(a,b,c,d,x \right) &= \int_0^x \int_0^\infty
\frac{t\theta e^{-b t - d t \theta }}{a + c \theta} d\theta dt.
\label{eq:zf:double_integral3}
\end{align}
Each double integral can be evaluated using standard methods in
terms of a sum of exponential integral functions as shown in
\dref{eq:zf:I1:solution}-\dref{eq:zf:I3:solution}, where $E_1(x) =
\int_x^\infty \frac{e^{-t}}{t} dt$. An outline of the solutions of
\dref{eq:zf:double_integral1}-\dref{eq:zf:double_integral3} is given
as follows.
$\tilde{I}_1$ in \dref{eq:zf:double_integral1} can be solved by
integrating over $t$ first and making use of the two identities in
\cite{GradRzy00}
\begin{align}
\int_0^\infty \frac{e^{- \lambda \theta}}{\alpha + \beta \theta} d
\theta = \frac{e^{\frac{\lambda \alpha}{\beta} }}{\beta} E_1 \left (
\frac{\lambda \alpha}{\beta} \right ),
\label{eq:zf:integral_identity1}
\end{align}
\begin{align}
\int_0^\infty \frac{dx}{\left(x + \alpha \right)\left(x + \beta
\right)} = \frac {\ln \left( \beta / \alpha \right ) }{ \beta
-\alpha },
\end{align}
to solve the resulting integral over $\theta$.
%
%
Next, we note that $\tilde{I}_2$ follows directly from $\tilde{I}_1$
as $\tilde{I}_2 = - \frac{\partial \tilde{I}_1}{\partial a}$. Hence,
by differentiating the expression for $\tilde{I}_1$ in
\dref{eq:zf:I1:solution}, we obtain \dref{eq:zf:I2:solution}. In
order to differentiate the exponential integral, we use Leibnitz's
integration formula to give
\begin{align}
\frac{\partial\left[ E_1 \left ( \alpha a \right ) \right]}{\partial
a}  = - \frac{e^{-\alpha a}}{a}.
\end{align}
$\tilde{I}_3$ in \dref{eq:zf:double_integral3} can also be solved by
a similar approach as in $\tilde{I}_1$ and making use of the
following integral identity from \cite{GradRzy00} where necessary:
\begin{align}
\int_0^x \frac{e^{-bt}}{t + d} dt = e^{bd} \left[ E_1 ( bd) - E_1 (
bx + bd ) \right].
\end{align}
%
%
%
%

%
%
%
%
\newpage
\begin{IEEEbiography}{Dushyantha Basnayaka}
(S'11) was born in 1982 in Colombo, Sri Lanka. He received the
B.Sc.Eng degree with 1\textsuperscript{st} class honors from the
University of Peradeniya, Sri Lanka, in Jan 2006. He is currently
working towards for his PhD degree in Electrical and Computer
Engineering at the University of Canterbury, Christchurch, New
Zealand.\\
He was an instructor in the Department of Electrical and Electronics
Engineering at the University of Peradeniya from Jan 2006 to May
2006. He was a system engineer at MillenniumIT (a member company of
London Stock Exchange group) from May 2006 to Jun 2009. Since Jun.
2009 he is with the communication research group at the University
of Canterbury, New Zealand.\\
D. A. Basnayaka is a recipient of University of Canterbury
International Doctoral Scholarship for his doctoral studies at UC.
His current research interest includes all the areas of digital
communication, specially macrodiversity wireless systems. He holds
one pending US patent as a result of his doctoral studies at UC.
\end{IEEEbiography}
\begin{IEEEbiography}{Peter Smith}
(M'93-SM'01) received the B.Sc degree in Mathematics and the Ph.D
degree in Statistics from the University of London, London, U.K., in
1983 and 1988, respectively. From 1983 to 1986 he was with the
Telecommunications Laboratories at GEC Hirst Research Centre. From
1988 to 2001 he was a lecturer in statistics at Victoria University,
Wellington, New Zealand. Since 2001 he has been a Senior Lecturer
and Associate Professor in Electrical and Computer Engineering at
the University of Canterbury in New Zealand. His research interests
include the statistical aspects of design, modeling and analysis for
communication systems, especially antenna arrays, MIMO, cognitive
radio and relays.
\end{IEEEbiography}
\begin{IEEEbiography}{Philippa Martin}
(S'95-M'01-SM'06) received the B.E. (Hons. 1) and Ph.D. degrees in
electrical and electronic engineering from the University of
Canterbury, Christchurch, New Zealand, in 1997 and 2001,
respectively.  From 2001 to 2004, she was a postdoctoral fellow,
funded in part by the New Zealand Foundation for Research, Science
and Technology (FRST), in the Department of Electrical and Computer
Engineering at the University of Canterbury.  In 2002, she spent 5
months as a visiting researcher in the Department of Electrical
Engineering at the University of Hawaii at Manoa, Honolulu, Hawaii,
USA.  Since 2004 she has been working at the University of
Canterbury as a lecturer and then as a senior lecturer (since 2007).
In 2007, she was awarded the University of Canterbury, College of
Engineering young researcher award.  She served as an Editor for the
IEEE Transactions on Wireless Communications 2005-2008 and regularly
serves on technical program committees for IEEE conferences.  Her
current research interests include multilevel coding, error
correction coding, iterative decoding and equalization, space-time
coding and detection, cognitive radio and cooperative communications
in particular for wireless communications
\end{IEEEbiography}


\begin{thebibliography}{1}



\bibitem{Wang10}
 L. C. Wang and C. J. Yeh, ``A three-cell coordinated network MIMO with fractional frequency reuse and directional antennas,''
 \emph{{IEEE} ICC}, Cape Town, South Africa, pp. 1--5, Mar. 2010.

 \bibitem{Yu09}
 H. Yu, S. Zhang, V. K. N. Lau and X. Yang, ``Game theoretical power control for
 open-loop network MIMO systems with partial cooperation,''
 \emph{{IEEE} TENCON}, Singapore, pp. 1--6, Oct. 2009.

\bibitem{Siva07}
 S. Venkatesan, A. Lozano and R. Valenzuela, ``Network MIMO: Overcoming intercell interference in indoor wireless systems,''
 \emph{{IEEE} ACSSC}, Pacific Grove, California, pp. 83--87, Jul. 2007.

\bibitem{Sander09}
 A. Sanderovich, O. Somekh, H.V. Poor and S. Shamai, ``Uplink macro diversity of limited backhaul cellular network,''
 \emph{{IEEE} Trans. Inform. Theory}, vol 55, no. 8, pp. 3457--3478, Aug.
 2009.

  \bibitem{Big07}
 E. Biglieri, R. Calderbank, A. Constantinides, A. Goldsmith, A. Paulraj and H. V. Poor, \emph{MIMO Wireless Communication},
 1st~ed, Cambridge: Cambridge University Press, 2007.


\bibitem{Wy94}
 A. Wyner, ``Shannon theoretic approach to a Gaussian cellular multiple access channel,''
 \emph{{IEEE} Trans. Inform. Theory}, vol 40, no. 6, pp. 1713--1727,
 Nov. 1994.

\bibitem{SaKiMo10}
 M. Sawahashi, Y. Kishiyama, A. Moromoto, D. Nishikawa and M. Tanno, ``Coordinated multipoint
 transmission/reception techniques for LTE-Advanced,''
\emph{{IEEE} Trans. Wireless Commun.}, vol. 17, no. 3, pp. 26--34,
Nov. 2011.


  \bibitem{Werner06}
 W. Weichselberger, M. Herdin, H. Ozcelik, and E. Bonek, ``A stochastic MIMO channel model with joint correlation of both link ends,''
 \emph{{IEEE} Trans. Wireless Commun}, vol. 5, no. 1, pp. 90--100, Jan 2006.

  \bibitem{Mari09}
 M. Kobayashi, M. Debbah and J. C. Belfiore, ``Outage efficient strategies for network MIMO with partial CSIT,''
 \emph{{IEEE} ISIT} 2009, Coex, Seoul, Korea, pp. 249--253, Jun/Jul. 2009.

\bibitem{Lin09}
 L. Hui and W. W. Bo, ``Performance analysis of network MIMO technology,''
 \emph{{IEEE} APCC}, Shanghai, China, pp. 234--236, May 2009.


  \bibitem{Gao98}
 H. Gao, P. J. Smith and M. V. Clark, ``Theoretical reliability of MMSE linear diversity
 combining in Rayleigh-fading additive interference channels,''
 \emph{{IEEE} Trans. Commun.}, vol.~46, no. 5, pp. 666--672, May. 1998.

\bibitem{Peter07}
 P. J. Smith, ``Exact performance analysis of optimum combining with multiple interferers
in flat Rayleigh fading,'' \emph{{IEEE} Trans. Commun.}, vol. 55,
no. 9, pp. 1674--1677, Sep. 2007.

  \bibitem{Winters94}
 J. H. Winters, J. Salz and R. D. Gitlin, ``The impact of antenna diversity on
 the capacity of wireless communication systems,'' \emph{{IEEE} Trans. Commun.}, vol. 42, no.2/3/4, pp.1740--1751, Feb/Mar/Apr. 1994.

%


  \bibitem{Gore02}
 D. A. Gore, R. W. Heath, Jr. and A. J. Paulraj, ``Transmit selection in spatial multiplexing systems,''
 \emph{{IEEE} Commun. Lett.}, vol.~6, no.11, pp. 491--493, Nov. 2002.

\bibitem{Jiang11}
 Y. Jiang, M. K. Varanasi, and J. Li, ``Performance analysis of ZF and
MMSE equalizers for MIMO systems: A closer study in high SNR
regime,'' \emph{{IEEE} Trans. Inform. Theory}, vol. 57, no. 4, pp.
2008 - 2026, Apr. 2011.


\bibitem{Salz94}
J. Salz and J. H. Winters, ``Effect of fading correlation on
adaptive arrays in digital mobile radio,'' \emph{{IEEE} Trans. Veh.
Technol.}, vol. 43, pp. 1049--1057, Nov. 1994.

\bibitem{Shiu00}
D. Shiu, G. Foschini, M. Gans, and J. Kahn, ``Fading correlation and
its effect on the capacity of multielement antenna systems,''
\emph{{IEEE} Trans. Commun.}, vo1.48, no.3, pp. 502--513, Mar. 2000.

\bibitem{Shin06}
H. Shin, M. Z. Win, J. H. Lee and M. Chiani, ``On the capacity of
doubly correlated MIMO channels,'' \emph{{IEEE} Trans. Wireless
Commun}, vol-5, no. 8, pp. 2253--2265, Aug 2006.

\bibitem{Mckay07}
M. R. McKay, A. Grant, and I. B. Collings, ``Performance analysis of
MIMO-MRC in double-correlated Rayleigh environments,'' \emph{{IEEE}
Trans. Commun.}, vol-55, no. 3, pp. 497--507, Mar 2007.


\bibitem{Muir82}
 R. J. Muirhead, \emph{Aspects of Multivariate Statistical Theory},  1st~ed, New York: John Wiley, 1982.

\bibitem{Gross89}
K. I. Gross and D. S. T. P. Richards, ``Total positivity, spherical
series and hypergeometric functions of matrix arguments,'' \emph{J.
Approx. Theory}, vol-59, pp. 224--226, 1989.


\bibitem{Winters84}
 J. H. Winters, ``Optimum combining in digital mobile radio with co-channel
interference,'' \emph{{IEEE} J. Select Areas in Commun.}, vol.SAC-2,
pp. 528--539, July. 1984.



\bibitem{Suwa72}
T. Suwa, ``Finite-sample properties of the k-class estimators,''
\emph{Econometrica}, vol. 40, no. 4, pp. 653--680, Jul 1972.

 \bibitem{GradRzy00}
 I. S. Gradshteyn and I. M. Ryzhik, \emph{Table of Integrals, Series, and Products},  6th~ed, Boston: Academic Press, 2000.

\bibitem{Proakis01}
 J. G. Proakis, \emph{Digital Communications}, 4th~ed, New York: McGraw-Hill, 2001.

\bibitem{Dush11}
 D. A. Basnayaka, P. J. Smith and P. A. Martin, ``Exact dual-user macrodiversity performance with linear receivers in flat Rayleigh
 fading,'' \emph{{IEEE} ICC}, Ottawa, Canada, pp.~ 5626--5631, Jun. 2012.

\bibitem{Lib94}
O. Lieberman, ``A Laplace approximation to the moments of a ratio of
quadratic forms,'' \emph{Biometrika}, vol. 81, no. 4, pp. 681--690,
Dec 1994.

 \bibitem{SiAl00}
M. K. Simon and M. S. Alouini, \emph{Digital Communications over
Fading Channels: A Unified Approach to Performance Analysis}.\hskip
1em plus 0.5em minus 0.4em\relax New York, NY, USA: Wiley, 2000.

 \bibitem{Gold00}
 A. Goldsmith, \emph{Wireless Communication},  1st~ed, Cambridge University Press, 2005.
\end{thebibliography}
\end{document}